# Cobalt-substitution-induced Superconductivity in a New Compound with ZrCuSiAs-type Structure, SrFeAsF

Satoru Matsuishi[1], Yasunori Inoue[2], Takatoshi Nomura[2], Masahiro Hirano[1,3] and Hideo Hosono[1,2,3]

[1] Frontier Research Center, Tokyo Institute of Technology, Mail Box S2-13, 4259 Nagatsuta, Midori-ku, Yokohama 226-8503, Japan

[2] Materials and Structures Laboratory, Tokyo Institute of Technology, Mail Box R3-1, 4259 Nagatsuta, Midori-ku, Yokohama 226-8503, Japan

[3] ERATO-SORST, JST, Frontier Research Center, Tokyo Institute of Technology, Mail Box S2-13, 4259 Nagatsuta, Midori-ku, Yokohama 226-8503, Japam.

We have synthesized a quaternary fluoroarsenide SrFeAsF with the ZrCuSiAs-type structure (P4/nmm, $a$ = 0.3999 and $c$ = 0.8973 nm), which is composed of an alternately stacked $(FeAs)^{\delta-}$ and $(SrF)^{\delta+}$ layers, analogous to the FeAs-based superconductor LaFeAsO. SrFeAsF shows metallic type conduction with the anomaly at ~180 K. The partial replacement of the Fe with Co suppresses the anomaly and induces the superconductivity, while the maximal $T_c$ (4 K for $SrFe_{0.875}Co_{0.125}AsF$) is much lower than that of the Co-substituted LaFeAsO. Replacement of $(LaO)^{\delta+}$ layers with $(SrF)^{\delta+}$ layers results in a enlargement of the $c$-axis length (+2.6%). These results suggest the importance of interlayer interaction as a critical $T_c$-controlling factor in FeAs-based superconductors.

The discovery of superconductivity in F-doped LaFeAsO with $T_c$ = 26 K[1] triggered intensive studies of FeAs-based layered compound systems including $R$FeAsO ($R$ = rare-earth) and $A$Fe$_2$As$_2$ ($A$ = alkali-earth) with a hope to realizing further high-$T_c$ superconductors.[2-11] These efforts lead to raising $T_c$ up to 56 K in Th-doped GdFeAsO so far.[8] The parent compounds for these superconductors belong to the ZrCuSiAs-type structure (space group P4/nmm) or ThCr$_2$Si-type structure (space group I4/mmm), consisting of an alternating stack of (FeAs)$^{\delta-}$ and ($RO$)$^{\delta+}$ or $A^{\delta+}$ layers. They suffer a crystallographic transition from the tetragonal to orthogonal, accompanying with antiferromagnetic spin ordering at 140-200 K.[1,3,8-14] Electron or hole doping to the (FeAs)$^{\delta-}$ layer through the doping of F$^-$ ions at the O sites of insulating layers in $R$FeAsO, for instance, (modulation doping) suppressed both transitions, resulting in emergence of the superconducting phase. Thus, the presence of the magnetic and/or charge instabilities is considered to be a key factor for the high-$T_c$. The efforts in the material studies have been focused on the synthesis of ZrCuSiAs-type and related-type compounds containing the square iron lattice[15-19] as well as on the carrier doping technique. Three successful doping methods have been reported to date, i.e., electron-doping by the oxygen-vacancy formation in insulating layers ($RO$)$^{\delta+}$ in $R$FeAsO,[20,21] hole-doping by alkali-metal doping to $A^{\delta+}$ layer in $A$Fe$_2$As$_2$[9-11] and electron-doping by partial replacement of Fe with Co in $R$FeAsO and $A$Fe$_2$As$_2$.[22-26] Effectiveness of the last technique in FeAs-based system is quite unique in comparison with high temperature cuprates.[27]

Here, we report the synthesis of a new FeAs containing ZrCuSiAs-type compound, SrFeAsF, in which the (FeAs)$^{\delta-}$ layer is sandwiched by (SrF)$^{\delta+}$ layer (inset of Fig 1a) and emergence of superconductivity by electron-doping to this material utilizing a partial substitution of Fe$^{2+}$ ions with a 3d$^6$ electronic configuration with Co$^{2+}$ ions with a 3d$^7$. It is noted that $T_c$ = 4 K is realized for the Co content of 12.5 % (SrFe$_{0.875}$Co$_{0.125}$AsF), which is much lower than $T_c$ in Co-substituted FeAs-based compound ($T_c$ = 14 K for LaFeAsO, 17 K for SmFeAsO, 20 K for SrFe$_2$As$_2$, and 22 K for BaFe$_2$As$_2$).[22-26] These findings suggest, (1) $T_c$ value is sensitive to the blocking layer, although FeAs layer seems to be essential for the high $T_c$ superconductivity, providing new information for governing factors for the $T_c$ values. (2) More generally, high $T_c$ superconductors still remain in a large number of ZrCuSiAs and related type crystals.[28]

Samples were prepared by a solid state reaction of $SrF_2$ (99.99 %), SrAs, $Fe_2As$ and $Co_2As$: $SrF_2$ + SrAs + (1−$x$) $Fe_2As$ + $x$ $Co_2As$ → $2SrFe_{1-x}Co_xAsF$. SrAs was synthesized by heating a mixture of Sr shots (99.99 wt. %) and As powders (99.9999 wt. %) at 650°C for 10 h in an evacuated silica tube. $Fe_2As$ and $Co_2As$ were synthesized from powders of respective elements at 800 °C for 10 h (Fe: 99.9 wt. %; Co: 99 wt. %). These products were then mixed in stoichiometric ratios, pressed, and heated in evacuated silica tubes at 1000 °C for 10 h to obtain sintered pellets. All the starting material preparation procedures were carried out in an Ar-filled glove box ($O_2$, $H_2O$ < 1 ppm).

The crystal structure and lattice constants of the materials were examined by powder X-ray diffraction (XRD; Bruker D8 Advance TXS) using Cu Kα radiation from a rotating anode with an aid of Rietveld refinement using Code TOPAS3.[29] Temperature dependence of DC electrical resistivity ($\rho$) at 2-300 K was measured by a four-probe technique using platinum electrodes deposited on samples. Magnetization ($M$) measurements were performed with a vibrating sample magnetometer (Quantum Design) in the same temperature range.

Figure 1 shows powder XRD patterns of undoped (a) and 12.5 atom% Co-substituted SrFeAsF (b). Almost all the peaks are assigned to those of the SrFeAsF phase, except several weak peaks arising from impurity phases ($SrF_2$ and FeAs, the volume fraction of these impurity phases being 3 % at most). The SrFeAsF phase is tetragonal symmetry with room-temperature lattice constants of $a$ = 0.3999 nm and $c$ = 0.8973 nm for the undoped sample and $a$ = 0.4002 nm and $c$ = 0.8943 nm for the 12.5 atom% Co-substituted sample (See Table I). The $c$-axis length of SrFeAsF is larger by 2.6 % than that of LaFeAsO, while $a$-axis length is smaller by 0.8 %.[12] In contrast to the replacement of La with other rare-earth ion (Ce, Pr, Nd, and Sm), the replacement of $(LaO)^{\delta+}$ layer with $(SrF)^{\delta+}$ layer leads to the shrinkage of $a$-axis length and the expansion of $c$-axis length. Upon substituting the Fe sites with Co, the $c$-axis length monotonically decreases with nominal Co concentration below 20 %, while the $a$-axis length remains almost constant (Fig.1b). The decrease in the $c$-axis indicates the enhanced Coulombic interaction between the $(SrF)^{\delta+}$ and $(FeAs)^{\delta-}$ layers with the Co content, providing evidence that the Co-substitution adds excess electrons to the $(FeAs)^{\delta-}$ layers.

Figure 2 (a) shows temperature ($T$) dependences of $\rho$ and molar magnetic susceptibility ($\chi_{mol}$) for undoped SrFeAsF. The $\chi_{mol}$-$T$ curve was obtained under the magnetic field ($H$) of 1 T with a zero field cooling mode. With a decrease in temperature, both $\rho$-$T$ and $\chi_{mol}$-$T$ curves exhibit sudden decreases at ~180 K ($T_{anom}$). This behavior is quite analogous to those of $R$FeAsO and $A$Fe$_2$As$_2$, indicating that the crystallographic transition accompanying with the magnetic order also occurs in SrFeAsF. As shown in Figure 2(b), 12.5% Co-doping apparently suppresses the anomaly, inducing superconductivity with the onset transition temperature ($T_{onset}$) of 4.8 K (See inset). $T_c$ defined as temperature where the $\rho$ value becomes half of that at $T_{onset}$ is 4 K. The sudden decrease in $\chi_{mol}$ due to the diamagnetism induced by the superconducting transition is also observed below $T_c$. Figure 2(c) shows volume magnetic susceptibility ($4\pi\chi$) vs. $T$ plots under zero-field cooling (ZFC) and field cooling (FC) with $H$ = 1 mT. For small magnetic field (< 1 mT), diamagnetic signal due to superconductivity is much larger than the paramagnetic signal, while it is masked by paramagnetic component in the high magnetic field region. Figure 2(d) shows the $M$-$H$ plot for SrFe$_{0.875}$Co$_{0.125}$AsF at 2 K, which demonstrate diamagnetic shielding below 1 mT. The volume fraction of the superconducting phase estimated from from the slope of $M$-$H$ curve in the region of 0-0.2 mT is ~ 17 %.

Figure 3(a) shows $\rho$-$T$ curves for Co-substituted SrFeAsF samples with several $x$ values. For $x$ = 0.05 and 0.075, the anomaly, observed in sample with $x$ = 0 at ~180 K, still appears as a small dull peak and the peak temperature shifts to lower temperatures with an increase in $x$. For $x$ = 0.1 and 0.125, the anomaly is completely suppressed and the superconducting transition is observed ($T_{onset}$ ~ 2 K for $x$ = 0.1). Further increase in the Co concentration above 15 % breaks the superconductivity and metallic conductivity ($d\rho/dT > 0$) is observed above $T_{min}$. The minimum ($T_{min}$) is observed to exist in the $\rho$-$T$ curves in the tetragonal phase for all Co concentrations. Figure 3(b) summarizes $T_{anom}$, $T_{onset}$ and $T_{min}$ as a function of $x$, demonstrating the Co-doping induces the superconducting phase in SrFeAsF and the highest $T_c$ of 4 K is attained at $x$ = 0.125.

Last, the present results are compared with those reported for other FeAs-based superconductors. It is worth noting that the threshold and optimal electron-doping levels are close to those of Co-substituted LaFeAsO and SmFeAsO although the $T_c$ value is fairly smaller. Lee et al. proposed an idea on the basis of many data reported to date that $T_c$ increases with the distortion of the FeAs$_4$ tetrahedra from the regular shape along the $c$-axis direction.[28] However, the angles in SrFeAsF ($\alpha$ = 111.1°) do not differ so largely from those of $R$FeAsO compounds (For example, $\alpha$ = 110.8° for SmFeAsO, $\alpha$ = 113.3° for LaFeAsO where $\alpha$ = 109.5° for regular tetrahedron),[14] indicating the distortion is not a dominant factor for the low $T_c$ in the Co-substituted SrFeAsF. The most prominent structural change observed by the replacement of (LaO)$^{\delta+}$ layers with (SrF)$^{\delta+}$ layers is a large anisotropic change in the $a$- and $c$-axes ($a$-axis:−0.8%, $c$-axis: +2.6%), in particular enlargement of the $c$-axis length. To date, no importance of the $c$-axis length, has been pointed out along with an experimental finding in the Fe-based layered superconductors.

In summary, the electrical conductivity and magnetization measurements demonstrate Co-substituted SrFeAsF is a bulk superconductor. $T_c$ changes with the Co-content, exhibiting the maximum of 4 K at the Co-content of ~12.5 atom %. Although the general feature of the phase diagram is similar to that of Co-substituted LaFeAsO in terms of the variation in $T_c$, and $T_{anom}$ with the Co content, $T_c$ is lowered significantly from LaO to SrF compounds. Crystallographic change in the FeAs layer cannot explain the large difference in $T_c$. These results suggest that expansion of the $c$-axis length yields a negative effect on $T_c$.

**Acknowledgment**

We thank Drs. Sung Wng Kim, Takashi Mine, Hiroshi Yanagi and Youichi Kamihara for their helpful discussions.

1) Y. Kamihara, T. Watanabe, M. Hirano, H. Hosono: J. Am. Chem. Soc. **130** (2008) 3296.
2) H. Takahashi, K. Igawa, K. Arii, Y. Kamihara, M. Hirano, H. Hosono: Nature **453** (2008) 376.
3) G. F. Chen, Z. Li, D. Wu, G. Li, W. Z. Hu, J. Dong, P. Zheng, J. L. Luo, N. L. Wang: Phys. Rev. Lett. **100** (2008) 247002.
4) Z-A. Ren, J. Yang, W. Lu, W. Yi, G-C. Che, X-L. Dong, L-L. Sun, Z-X. Zhao: Materials Research Innovations **12** (2008) 105.
5) Z-A. Ren, J. Yang, W. Lu, W. Yi, Z-L. Shen, Z-C. Li, G-C. Che, X-L. Dong, L-L. Sun, F. Zhou, Z-X. Zhao: Europhys. Lett. **82** (2008) 57002.
6) X. H. Chen, T. Wu, G. Wu, R. H. Liu, H. Chen, D. F. Fang, Nature **453** (2008) 761.
7) Z-A. Ren, W. Lu, J. Yang, W. Yi, X-L. Shen, Z-C. Li, G-C. Che, X-L. Dong; L-L. Sun, F. Zhou, Z-X. Zhao: Chin. Phys. Lett., **25** (2008) 2215.
8) C. Wang, L. Li, S. Chi, Z. Zhu, Z. Ren, Y. Li, Y. Wang, X. Lin, Y. Luo, S. Jiang, X. Xu, G. Cao, Z. Xu: Europhy. Lett. **83** (2008) 67006.
9) M. Rotter, M. Tegel, D. Johrendt: Phys. Rev. Lett. **101** (2008) 107006.
10) G. F. Chen, Z. Li, G. Li, W. Z. Hu, J. Dong, X. D. Zhang, P. Zheng, N. L. Wang, J. L. Luo: Chin. Phys. Lett. **25** (2008) 3403.
11) G. Wu, H. Chen, T. Wu, Y. L. Xie, Y. J. Yan, R. H. Liu, X. F. Wang, J. J. Ying, X. H. Chen: cond-mat/0806.4279.
12) C. de la Cruz, Q. Huang, J. W. Lynn, J. Li, W. Ratcliff II, J. L. Zarestky, H. A. Mook, G. F. Chen, J. L. Luo, N. L. Wang, P. Dai: Nature **453** (2008) 899.
13) T. Nomura, S. W. Kim, Y. Kamihara, M. Hirano, P. V. Sushko, K. Kato, M. Takata, A. L. Shluger, H. Hosono: cond-mat/0804.3569.
14) A. Martinelli, A. Palenzona, C. Ferdeghini, M. Putti, E. Emerich: cond-mat/0808.1024.
15) X. C. Wang, Q. Q. Liu, Y. X. Lv, W. B. Gao, L. X. Yang, R. C. Yu, F. Y. Li, C. Q. Jin: cond-mat/0806.4688.
16) F-C. Hsu, J-Y. Luo, K-W. Yeh, T-K. Chen, T-W. Huang, P. M. Wu, Y-C. Lee, Y-L. Huang, Y-Y. Chu, D-C. Yan, M-K. Wu: cond-mat/0807.2369.
17) Y. Mizuguchi, F. Tomioka, S. Tsuda, T. Yamaguchi, Y. Takano: cond-mat/0807.4315.
18) K-W. Yeh, T-W. Huang, Y-L. Huang, T-K. Chen, F-C. Hsu, P. M. Wu, Y-C. Lee, Y-Y.

Chu, C-L. Chen, J-Y. Luo, D-C. Yan, M-K. Wu: cond-mat/0808.0474.

19) H. Hosono: J. Phys. Soc. Jpn. submitted

20) Z. A. Ren, G. C. Che, X. L. Dong, J. Yang, W. Lu, W. Yi, X. L.Shen, Z. C. Li, L. L. Sun, F. Zhou, and Z. X. Zhao: Europhys. Lett. **83** (2008) 17002.

21) H. Kito, H. Eisaki, and A. Iyo: J. Phys. Soc. Jpn. **77** (2008) 063707.

22) A. S. Sefat, A. Huq, M. A. McGuire, R. Jin, B. C. Sales, D. Mandrus: Phys. Rev. B **78** (2008) 104505.

23) G. Cao, C. Wang, Z. Zhu, S. Jiang, Y. Luo, S. Chi, Z. Ren, Q. Tao, Y. Wang, Z. Xu: cond-mat/0807.1304.

24) Y. K. Li, X. Lin, Z. W. Zhu, H. Chen, C. Wang, L. J. Li, Y. K. Luo, M. He, Q. Tao, H. Y. Li, G. H. Cao, Z. A. Xu: cond-mat/0808.3254.

25) A. S. Sefat, R. Jin, M. A. McGuire, B. C. Sales, D. J. Singh, D. Mandrus: Phys. Rev. Lett. **101** (2008) 117004.

26) A. Leithe-Jasper, W. Schnelle, C. Geibel, H. Rosner: cond-mat/0807.2223.

27) J. M. Tarascon, L. H. Greene, P. Barboux, W. R. McKinnon, G. W. Hull: Phys. Rev. B. **36** (1987) 8393.

28) R. Pottgen, D. Johrendt: cond-mat/0807.2138.

29) TOPAS, Version 3; Bruker AXS: Karlsruhe Germany, (2005).

28) C. H. Lee, A. Iyo, H. Eisaki, H. Kito, M. T. Fernandez-Diaz, T. Ito, K. Kihou, H. Matsuhata, M. Braden, K. Yamada: J. Phys. Soc. Jpn. **77** (2008), 083704.

Figure 1 (a) Powder XRD patterns of SrFeAsF (+) and Rietveld fit (red line): Green line is difference between these patterns. Bars at bottom show calculated Bragg diffraction positions of SrFeAsF and $SrF_2$. Inset shows structure model of SrFeAsF. (b) XRD pattern of $SrFe_{0.875}Co_{0.125}AsF$. Inset shows lattice constants $a$ and $b$ as a function of Co content. Bars at bottom show calculated diffraction positions of $SrFe_{0.875}Co_{0.125}AsF$, $SrF_2$ and FeAs.

Figure 2 Electrical resistivity ($\rho$) and molar magnetic susceptibility ($\chi_{mol}$) vs. temperature ($T$) plots for undoped SrFeAsF (a) and $SrFe_{0.875}Co_{0.125}AsF$ (b). $H = 1$ T for $\chi_{mol}$-$T$ measurements. (c) Zero-field cooling (ZFC) and field cooling (FC) volume magnetic susceptibility ($4\pi\chi$) vs. $T$ plots for $SrFe_{0.875}Co_{0.125}AsF$ with $H = 1$ mT. (d) Magnetization ($M$) vs. Magnetic Field ($H$) plot for $SrFe_{0.875}Co_{0.125}AsF$. Volume fraction of superconducting phase is estimated from the slope of liner region of $M$-$H$ curve (dashed line).

Figure 3 (a) Electrical resistivity versus temperature plot for $SrFe_{1-x}Co_xAsF$: $x = 0.0$, 0.05, 0.075, 0.10, 0.125, 0.15 and 0.2. Horizontal line under each plot denotes $\rho / \rho_{300} = 0$ line. (b) $T_c$, $T_{onset}$ and $T_{min}$ in the $\rho$-$T$ curves as a function of $x$. $T_c$ is defined as the temperature where the value becomes half of that at $T_{onset}$. $T_{anom}$ values for $x = 0$, 0.05 and 0.75 are also shown.

Table I. Atomic parameters of SrFeAsF and SrFe$_{0.875}$Co$_{0.125}$AsF (space group P4/nmm) determined by Rietveld refinements of powder X-ray diffraction data at 300 K. $B_{eq}$ is the isotropic atomic displacement parameter.

(a) SrFeAsF

$a$ = 3.999383(28) Å, $c$ = 8.97274(10) Å

$R_{wp}$ = 13. 76 %, $S$ = 1.98

| Atom | site | occ. | x | y | z | $B_{eq}$ (Å$^2$) |
|---|---|---|---|---|---|---|
| Sr | 2c | 1 | 1/4 | 1/4 | 0.15903(13) | 0.487(33) |
| Fe | 2b | 1 | 3/4 | 1/4 | 1/2 | 0.191(36) |
| As | 2c | 1 | 1/4 | 1/4 | 0.65280(18) | 0.166(35) |
| F | 2a | 1 | 3/4 | 1/4 | 0 | 0.55(12) |

(b) SrFe$_{0.875}$Co$_{0.125}$AsF

$a$ = 4.001816(23) Å, $c$ = 8.943446(76) Å

$R_{wp}$ = 10.21 %, $S$ = 1.78

| Sr | 2c | 1 | 1/4 | 1/4 | 0.15821(12) | 0.383(40) |
|---|---|---|---|---|---|---|
| Fe | 2b | 7/8 | 3/4 | 1/4 | 1/2 | 0.243(39) |
| Co | 2b | 1/8 | 3/4 | 1/4 | 1/2 | 0.243(39) |
| As | 2c | 1 | 1/4 | 1/4 | 0.65095(17) | 0.130(40) |
| F | 2a | 1 | 3/4 | 1/4 | 0 | 0.72(13) |

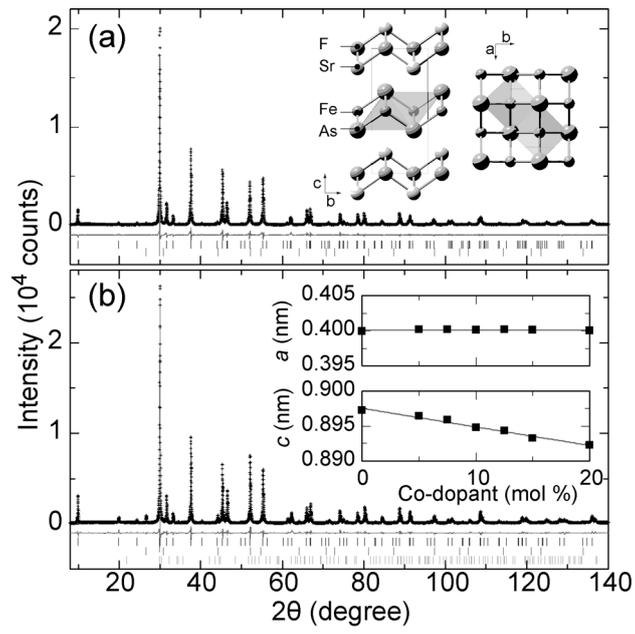

Fig.1

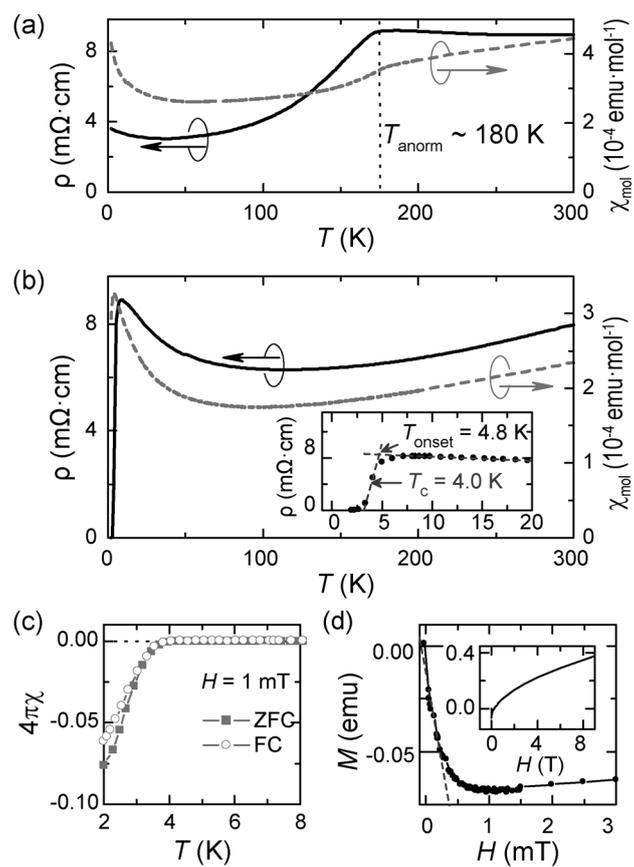

Fig. 2

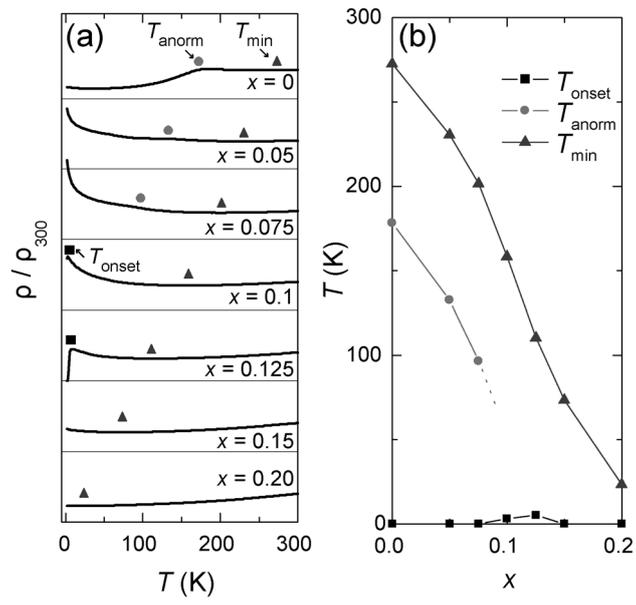

Fig. 3